\documentclass[twocolumn]{aastex62}

\usepackage{textcomp}
\usepackage{amsmath}

\submitjournal{The Planetary Science Journal}

\begin{document}

\title{Composition and Size Dependent Sorting in Preplanetary Growth: Seeding the Formation of Mercury-like Planets}

\correspondingauthor{Maximilian Kruss}
\email{maximilian.kruss@uni-due.de}

\author{Maximilian Kruss}
\author{Gerhard Wurm}

\affiliation{University of Duisburg-Essen}

\begin{abstract}
In an earlier work, we found that large metallic iron fractions in dust aggregates and strong magnetic fields boost preplanetary growth. This sets an initial bias for the formation of Mercury-like planets in the inner part of protoplanetary disks. We extended these experiments here by adding pure quartz aggregates to the iron-rich aggregates. Magnetic boost still leads to the formation of larger clusters of aggregates. These clusters now include silicate aggregates, which can also be connecting bridges between chains. However, at least a certain fraction of iron-rich aggregates are needed to trigger magnetic boost. Without a magnetic field, the sticking properties of the aggregates and their constituents determine the composition of clusters of a given size. This introduces a new fractionation and sorting mechanism by cluster formation at the bouncing barrier. 
\end{abstract}

\section{Introduction}

As in \citet{Kruss2018}, the motivation behind this work is still the high iron fraction of some rocky planets. This includes Mercury but also iron-rich exoplanets orbiting close to their host stars \citep{Spohn2001, Hauck2013, Rappaport2013, Guenther2017, Sinukoff2017, Santerne2018, Margot2019, Price2019}.

Quite a number of processes have been proposed to explain higher iron ratios than expected, including evaporation \citep{Cameron1985}, large impacts \citep{Benz1988, Stewart2013, Asphaug2014}, photophoresis \citep{Wurm2013, Cuello2016}, inward drift of interplanetary dust particles \citep{Ebel2011}, and magnetic erosion \citep{Hubbard2014}. In \citet{Kruss2018} (hereafter Paper I), we introduced a new mechanism that we will call magnetic boost here. Boost refers to an increase in the maximum size of particles at the bouncing barrier in planetesimal formation \citep{Zsom2010, Kelling2014, Kruss2016, Kruss2017}.

While aggregation of magnets in the context of planetesimal formation has been studied before by \citet{Nuth1994}, \citet{Dominik2002}, and \citet{Nuebold2003}, the boost reported here and in Paper I is related to magnetization due to magnetic fields in protoplanetary disks. These fields can be up to several mT at the inner disk edge and decrease with radial distance \citep{Donati2005, Wardle2007, Dudorov2014, Bertrang2017, Brauer2017, Maurel2019}. The distribution of iron in the disk is expected to follow a similar trend as, in the solar system, the iron content decreases with radial distance from the Sun \citep{Trieloff2006}. The magnetic fields magnetize iron grains, leading to the formation of large chain-like clusters of iron-rich aggregates. Such clusters can be several times larger than the size of individual aggregates at the bouncing barrier \citep{Kruss2018}. The idea of setting a bias for Mercury-like planets is that larger clusters might be more prone to drag instabilities and planetesimal formation \citep{Youdin2005, Bai2010, Drazkowska2014, Johansen2014, Simon2016}. 

Our first experiments in Paper I, where only iron-rich aggregates were used, clearly showed the potential of magnetic boost. Here, we approach more realistic conditions as the protoplanetary disk contains more than just one dust species. The abundances of iron and silicate in the solar system are similar \citep{Palme2014}. In addition, a significant amount of iron is incorporated in silicates with olivine and pyroxene being the most common minerals \citep{Boekel2004, Zhukovska2018}. Specifically, we extend the previous experiments by not only using iron-rich aggregates but by adding pure nonmagnetic silicate (quartz) aggregates.

\section{Experiment}

The focus of this work is the evolution of levitated aggregates composed of iron and quartz under different conditions. To judge the collisional outcome and potential cluster growth, it is important to know the sticking properties of the used samples, which are discussed in section \ref{sec.tensile} while section \ref{sec.lev} introduces the levitation setup.

\subsection{\label{sec.tensile} Tensile Strength Measurement}

The surface energy is a key parameter concerning the outcome of a collision. It influences the threshold velocity for sticking and therefore also the size of grown clusters \citep{Dominik1997, Guettler2010, Wada2013}. Two different dust species and mixtures of both are treated in the experiments. We used quartz and iron grains with average grain diameters of 3.0\,\textmu m and 2.2\,\textmu m and densities of 2.65\,g/cm$^3$ and 7.87\,g/cm$^3$, respectively. The shape of the grains is an important parameter that may influence the stability of aggregates. While the quartz grains are known to be nonspherical, the exact shape of the iron grains remains uncertain. For detailed size distributions and magnetic properties of the iron particles, see Paper I.

\begin{figure}[h]
	\begin{centering}
		\includegraphics[width=0.6\columnwidth]{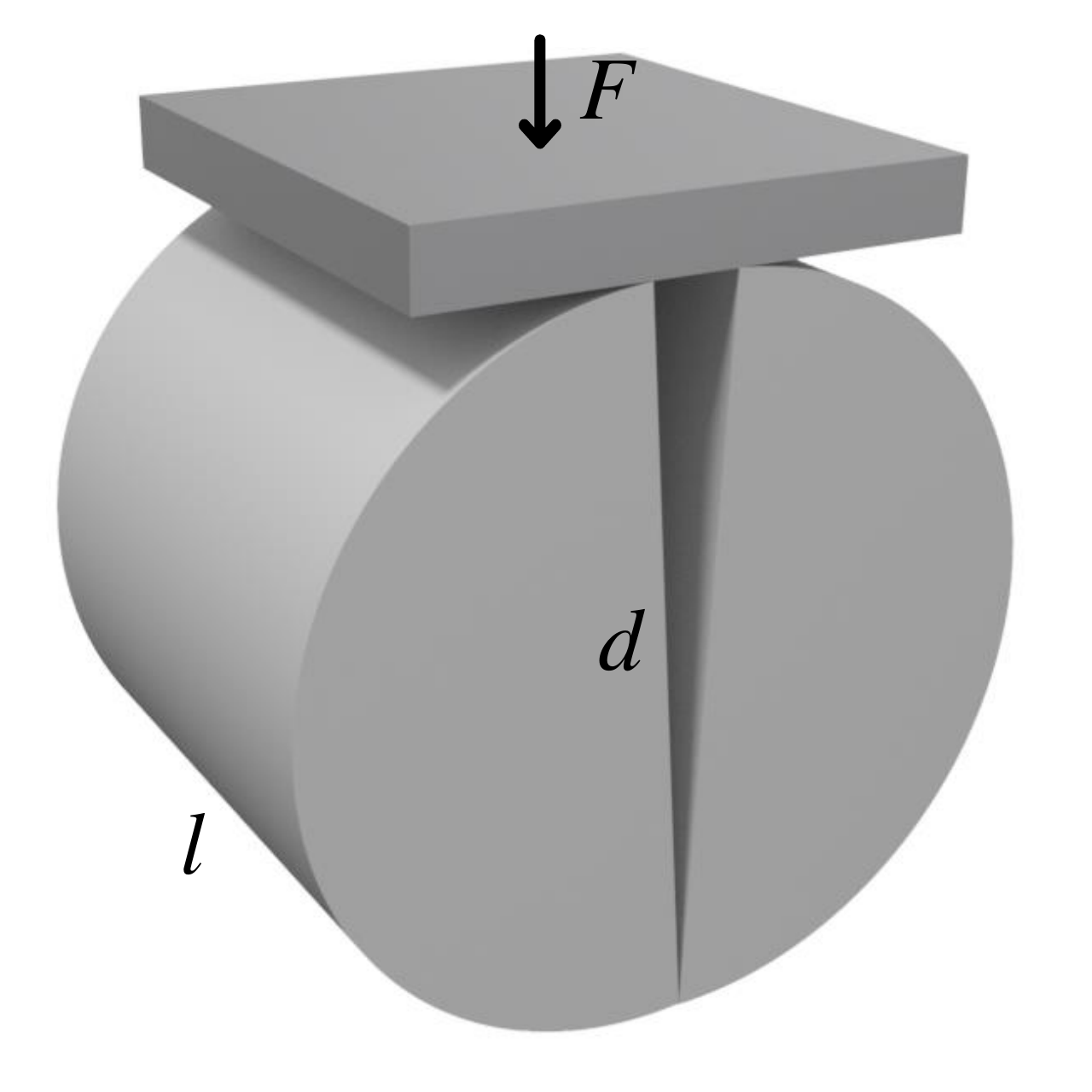}
		\caption{\label{fig.brazil} Principle of the Brazilian test. A cylinder of pressed dust with diameter $d$ is split in two by exerting a force, $F$, onto the mantle face with length $l$. Typical values are $d=l=8$\,mm and $F_{\rm{split}}=1$\,N.}
	\end{centering}
\end{figure}

The levitation setup in section~\ref{sec.lev} does not allow a quantitative measurement of sticking properties. A more suitable way is using the Brazilian test, which is sketched in Figure~\ref{fig.brazil}. This method is based on measuring the force needed to split a cylinder of pressed dust in two equal parts. For details of the experimental procedure, we refer to \citet{Steinpilz2019a}. The splitting force $F_{\rm{split}}$ can be translated into the tensile strength $\sigma$ of the sample according to $\sigma=2F_{\rm{split}}/\pi dl$, where $d$ and $l$ are the cylinder's diameter and length, respectively.

\begin{figure}[h]
	\includegraphics[width=\columnwidth]{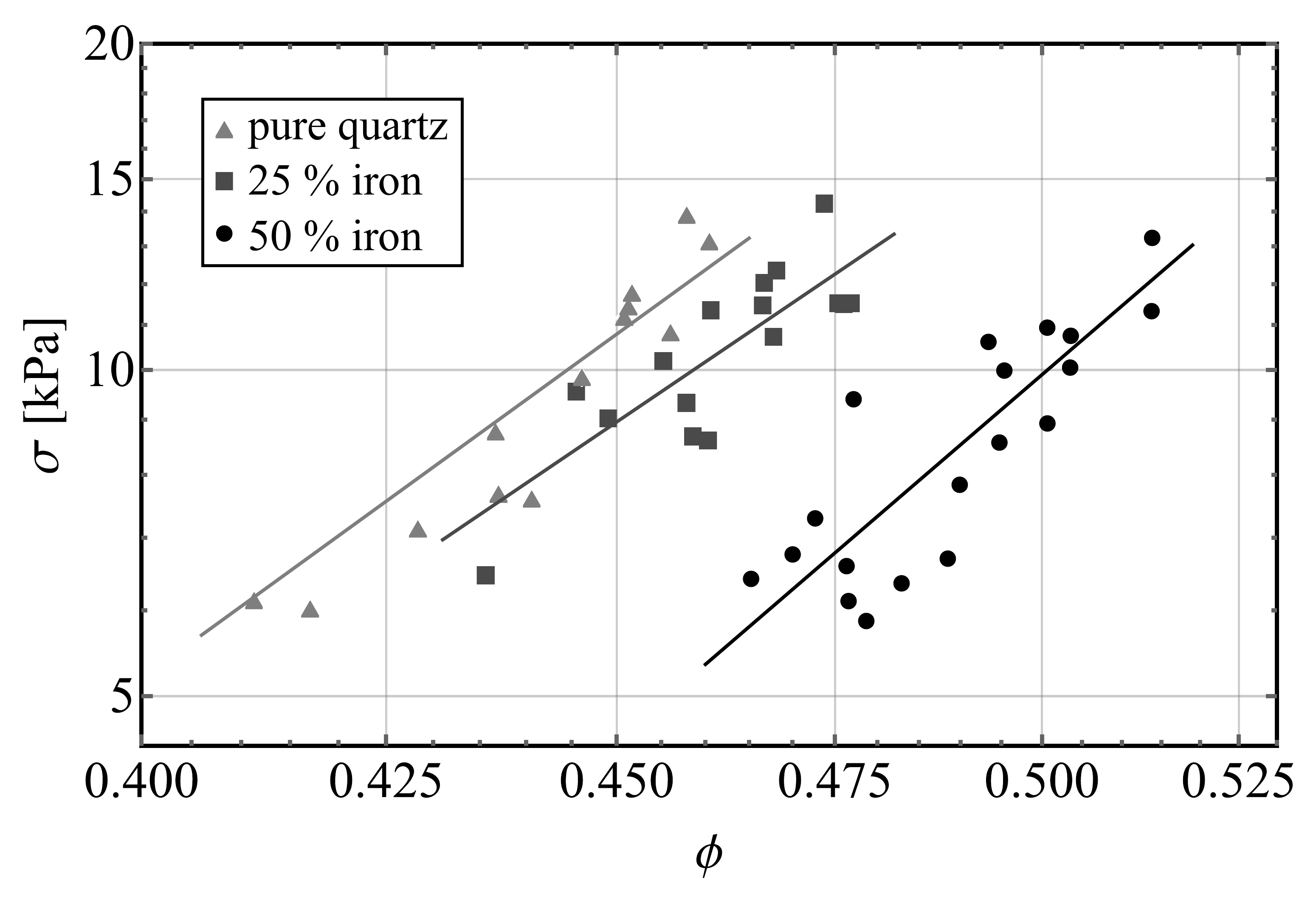}
	\caption{\label{fig.tensile} Tensile strength $\sigma$ over volume filling factor $\phi$ for three samples with different iron mass content. The straight lines suggest power-law dependences.}
\end{figure}

Figure~\ref{fig.tensile} shows the tensile strengths of three different samples prepared with iron contents of 0\,\%, 25\,\% and 50\,\% by mass. Depending on the preparation, the cylinders have different volume fillings, which can be quantified by the ratio of the measured cylinder density and the bulk density of the material: $\phi=\rho_{\rm{cyl}}/\rho_{\rm{bulk}}$. The tensile strength can be expressed using Rumpf's equation \citep{Rumpf1970}:
	\begin{equation}
		\label{eq.rumpf}
		\sigma=\frac{9\phi N}{8\pi d^2}F,
	\end{equation}
where $N$ is the number of neighbors in contact and $d$ is the particles' diameter. The force $F$ needed to break a contact is proportional to the surface energy $\gamma$ or
\begin{equation}
\label{jkr}
F = 3 \pi \gamma R     
\end{equation}
where $R$ is the reduced radius, assuming two spherical grains of same material collide \citep{Johnson1971, Dominik1997}. Considering not only the material property but also the shape and geometry of the sample, sticking can be described by the concept of an effective surface energy $\gamma_{\rm{eff}}$. The data in Figure~\ref{fig.tensile} show a decrease of the tensile strength with increasing iron content. Considering equation~\ref{eq.rumpf} and \ref{jkr}, this trend implies a decrease of $\gamma_{\rm{eff}}$ as well. For the iron-rich sample, $\gamma_{\rm{eff}}$ drops by a factor of more than 2 compared to pure quartz.

The exact relation between composition and surface energy is not the primary focus of this work, though it is notable that the effective surface energy drops the more iron the sample contains.

\subsection{\label{sec.lev} Levitation Setup}

\begin{figure}[h]
	\includegraphics[width=\columnwidth]{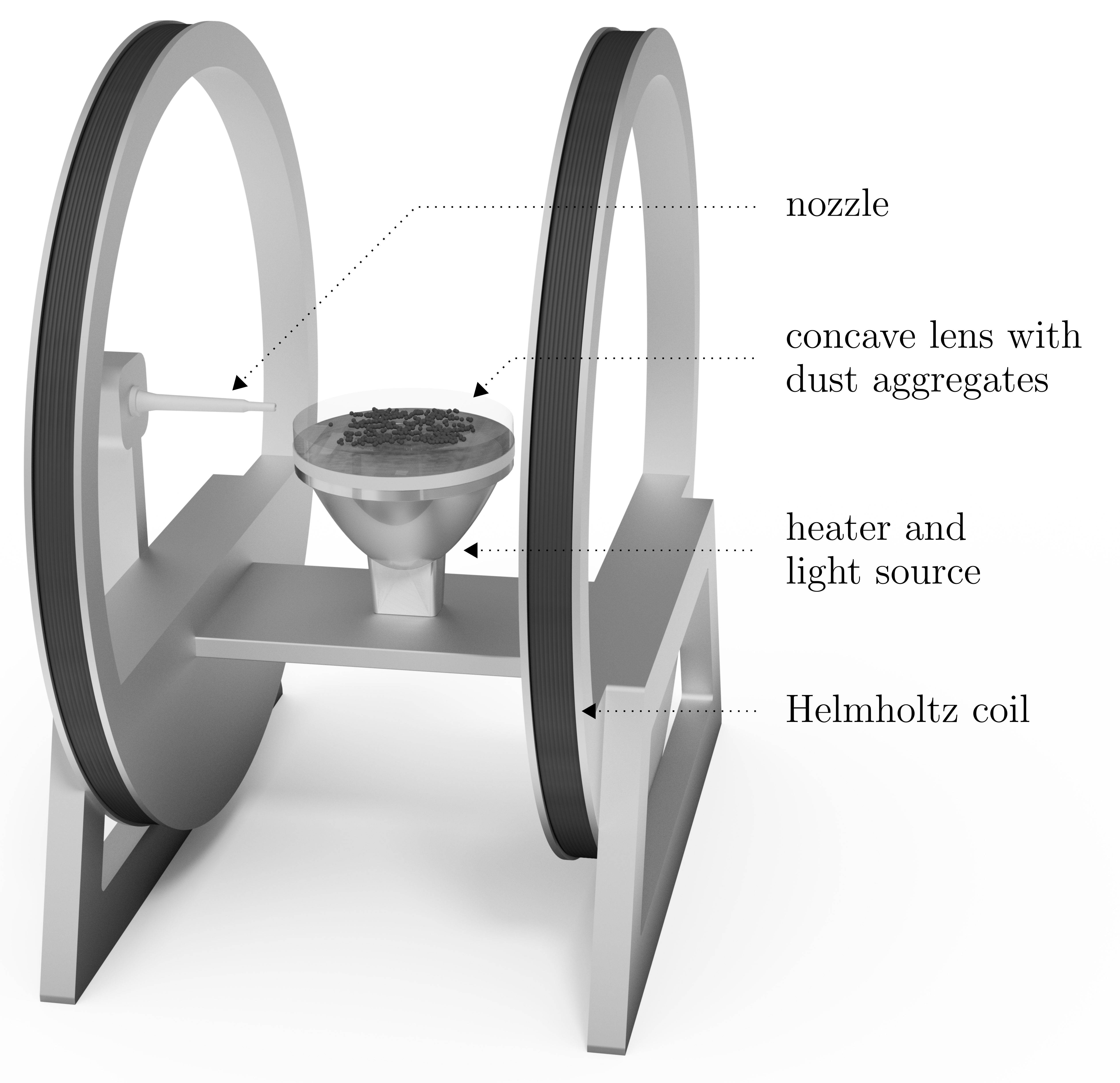}
	\caption{\label{fig.setup} Dust is levitated over a hot surface at low pressure. Grains collide and grow to a maximum size. A magnetic field can be applied (from \citet{Kruss2018}).}
\end{figure}

For readability, we show the sketch of the experiment from Paper I again in Figure~\ref{fig.setup}. Dust aggregates are placed on a heater. At low pressure, they are levitated by means of thermal creep \citep{Kelling2014}. Aggregates cannot slide over each other as their levitation height is comparable. They are then free to move in 2D and to collide with each other. Some injection of gas at the side of the levitator increases the velocity of the aggregates. Typical collision speeds are mm\,s$^{-1}$ to several cm\,s$^{-1}$ \citep{Kruss2016, Kruss2017}. The sample is observed from above by a camera at a frame rate of 100\,fps and a spatial resolution of 25\,\textmu m. This frame rate ensures that the formation of clusters and their survival over several frames can be observed in the highly dynamic experiments. In contrast to Paper I, the sample is illuminated from above by a ring light to distinguish between different aggregate species. Once the aggregates are levitated, a magnetic field can be applied to study the influence of magnetic forces. For that purpose, the levitator is placed in the central part of two Helmholtz coils, which generate a homogeneous magnetic field of up to 7\,mT. This value fits into the range of estimated field strengths in the inner regions of protoplanetary disks (e.g. \citet{Donati2005, Wardle2007}).

To avoid confusion, we note that we have three different size scales in the experiment going by different names here.

\begin{figure}[h]
	\includegraphics[width=\columnwidth]{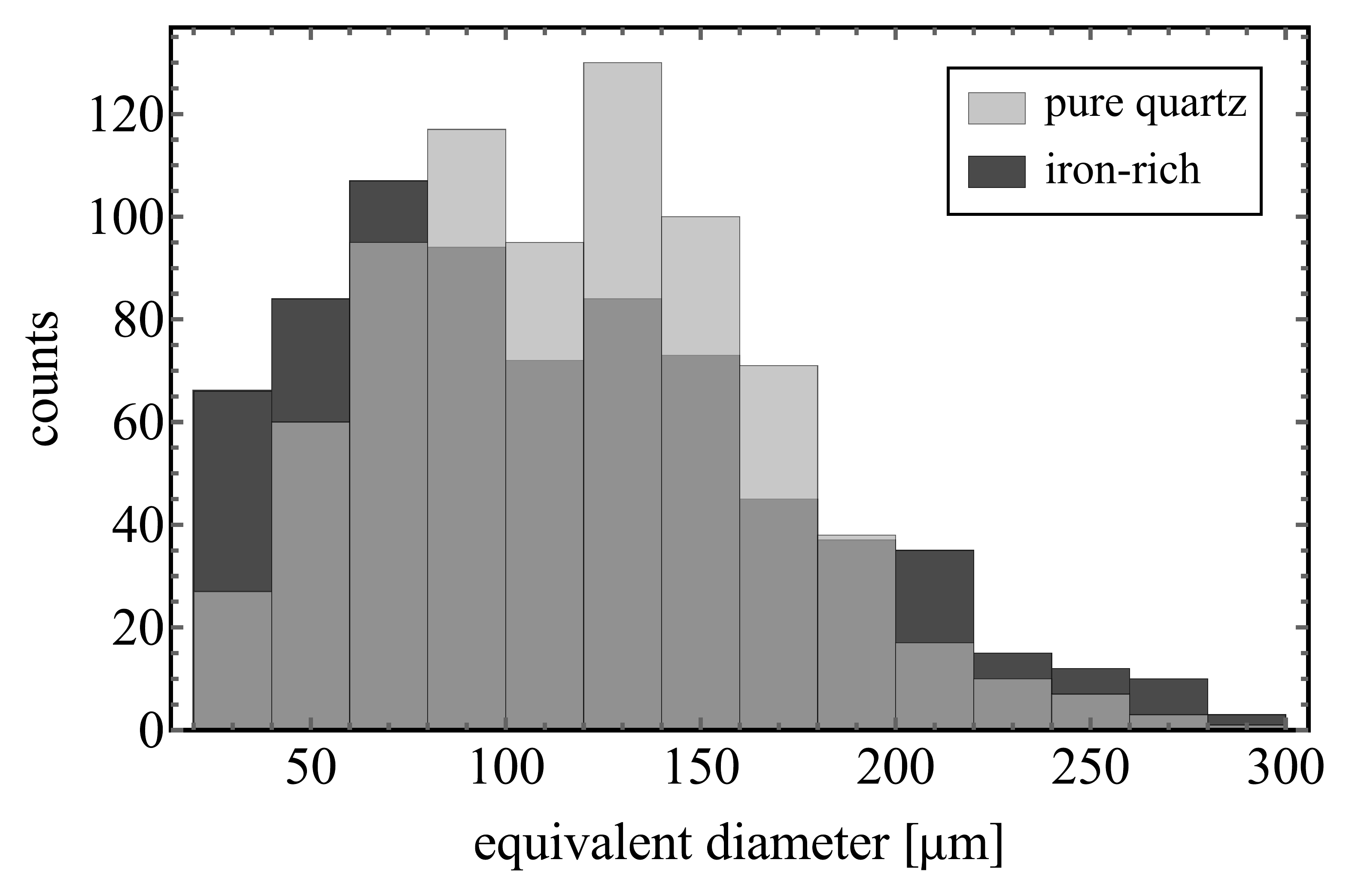}
	\caption{\label{fig.initials} Size distribution of initial aggregates after sieving through a 180\,\textmu m mesh. The equivalent diameter is calculated from the cross section $A$ according to $\sqrt{4A/\pi}$.}
\end{figure}

\begin{figure}[h]
	\includegraphics[width=\columnwidth]{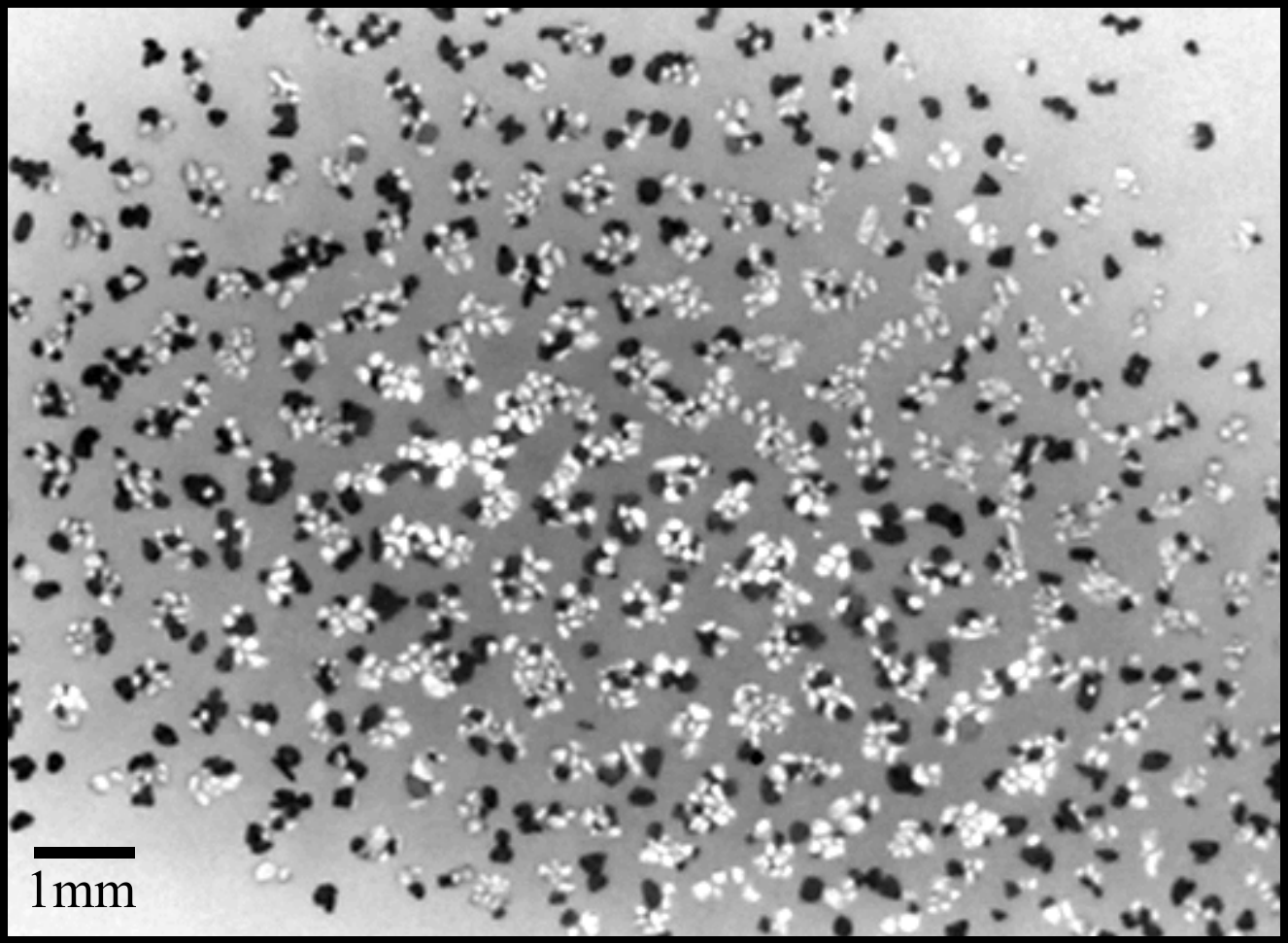}
	\caption{\label{fig.example1} Snapshot of clusters without magnetic field with an area fraction of iron-rich aggregates (dark) of 40\,\%. Quartz aggregates appear bright. The figure only shows a section of the full image.}
\end{figure}

\begin{itemize}
    \item Dust. These are the basic solid particles of micrometer size. We used quartz and iron grains with average grain diameters of 3.0\,\textmu m and 2.2\,\textmu m and densities of 2.65\,g/cm$^3$ and 7.87\,g/cm$^3$, respectively. These grains are not directly visible in the experiments due to a lower spatial resolution.
    
    \item Aggregate. Aggregates of dust grains that stick together already form in the sample container, no matter if they are iron-rich or quartz. For iron-rich aggregates, iron and quartz grains were premixed before aggregates were formed. Unless otherwise indicated, the mass ratio of iron and quartz in the aggregates was 1:1. As seen in Paper I, this results in homogeneous mixtures for these aggregates. We note that in this case, due to the difference in the density of the two constituents, the volume fractions are not equal. An aggregate consists of around 75\,\% quartz and 25\,\% iron grains. Pure silicate (quartz) aggregates are prepared from a pure quartz sample. The aggregates are sieved through a standard mesh of diameter $d_0$. For most experiments, we used a 180\,\textmu m mesh. The size distribution of the initial aggregates after sieving is shown in Figure~\ref{fig.initials}. In both cases, quartz and iron-rich, the size of the mesh limits the maximum size of aggregates. A few aggregates with an equivalent diameter larger than the mesh size are able to pass the sieve due to their uneven shapes or they accidentally drop on the same spot. Typically, the initial aggregates occupy around 10\,\% - 20\,\% of the monolayer area.
	
    \item Cluster. The size of the aggregates is only determined by the mesh and does not relate to the maximum size to which a dust sample might evolve in a self-consistent evolution. Therefore, aggregates, once free to move and collide, stick to each other and, this way, grow further. In these grown clusters, the individual aggregates can still be recognized. As the aggregates only move in one layer, clusters that are evolving from sticking collisions are restricted to 2D as well. The thickness of the clusters is determined by the initial aggregate size, which is around 180\,\textmu m in most experiments. The constituents in clusters mixed of iron-rich and quartz aggregates can be distinguished by their gray scale (see Figure~\ref{fig.example1}). With the cross sections of aggregates composing the respective cluster, we define the area fraction of iron-rich material $f_{\rm{ir}}$ within the cluster.  
\end{itemize}

It has to be noted that the sieving procedure rather involves a single vibration than a long-term shaking of the whole sample since this would induce a significant amount of electrical charge on the aggregates due to tribocharging \citep{Jungmann2018, Wurm2019}. Considerable attractive forces on approaching aggregates could be a consequence. In fact, there are a number of studies on the influence of electrical charges on planet formation, e.g. \citet{Steinpilz2019b, Steinpilz2020}. Figure~\ref{fig.initials} indicates similar size distributions for iron-rich and quartz aggregates after sieving. No large cluster is formed due to potential electrical charges prior to the experiment. However, this does not rule out any collisional charging of the aggregates once they are levitated, but this experiment is not suitable to measure this. We also did not see any effect of the magnetic field on potential static charges, even when applying the maximum field of 7\,mT to pure quartz aggregates only. All in all, we did not notice any evidence of charge-induced behavior in our experiments.

\begin{figure}[h]
	\includegraphics[width=\columnwidth]{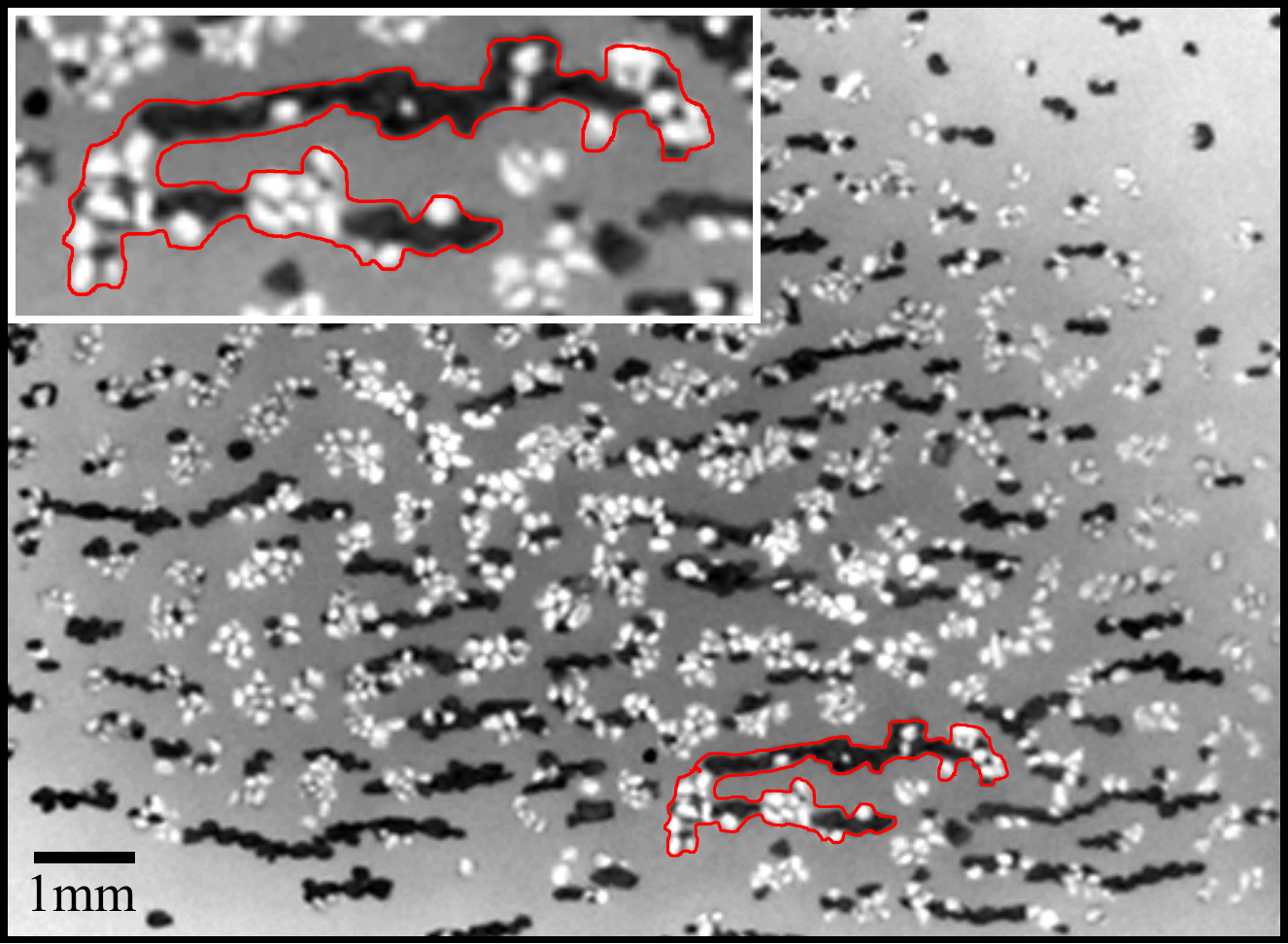}
	\caption{\label{fig.example2} Snapshot of clusters within a magnetic field of 7\,mT  and with an area fraction of iron-rich aggregates of 40\,\%. An example of a large cluster with quartz aggregates connecting different linear chains of iron-rich aggregates is marked. The figure only shows a section of the full image.}
\end{figure}

An example of the clusters formed without magnetic field is shown in Figure~\ref{fig.example1}. Depending on the initial aggregate size, the excitation by the gas flow, and the magnetic field, aggregates grow to larger clusters. Their 2D sizes as well as the cross section of iron-rich aggregates incorporated in these clusters are analyzed manually from the taken images. A cluster is counted separately as long as it is not in contact with another cluster. Once the magnetic field is switched on, iron-rich aggregates form chain-like clusters as seen in Figure~\ref{fig.example2}.

In general, the experiment shows both the growth and the deformation of clusters as the system is highly dynamic. Clusters are rearranged within much less than a second and no steady state is reached, no matter if the magnetic field is applied or not. This represents the evolution of clusters in the protoplanetary disk which rearrange at the bouncing barrier. However, it has to be noted that collisional timescales in the disk are completely different. An experimental run that lasts for several seconds simulates the evolution over several thousand years in the disk.

\section{Results}

\subsection{Composition dependent growth}

\begin{figure}[h]
	\includegraphics[width=\columnwidth]{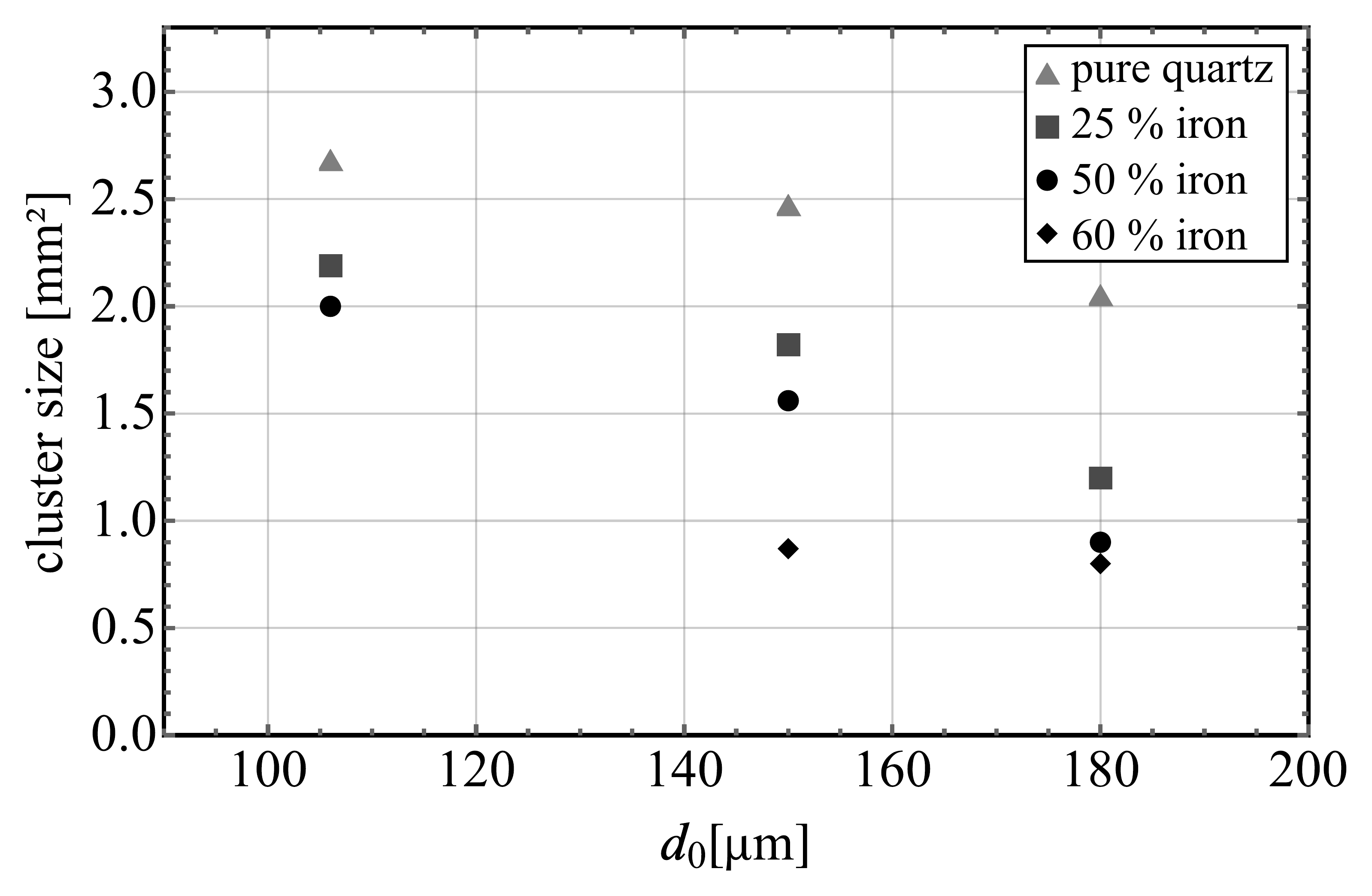}
	\caption{\label{fig.clustersize} Average cluster size depending on the sieve diameter $d_0$ and on the iron mass fraction within the aggregates.}
\end{figure}

While the characterization of the samples in section \ref{sec.tensile} is based on a static analysis, the levitation setup allows to study the collisional evolution of a large ensemble of aggregates. At first, only one sort of aggregate was used. The mass fraction of iron within the aggregates was varied for each experiment from no iron (pure quartz) to a mixture of 60\,\% iron and 40\,\% quartz. In addition, we varied the initial aggregate size by changing the sieve size $d_0$. Figure~\ref{fig.clustersize} shows the average 2D size of clusters that evolve depending on the initial aggregate size as well as on the iron mass fraction the aggregates were prepared with. We note that, here, only one species of aggregates is used per experiment and no magnetic field is applied. Aggregates collide and grow until they reach the bouncing barrier just like observed in the experiments by \citet{Kruss2017}.

Figure~\ref{fig.clustersize} clearly supports the trend revealed in the tensile strength measurement. Iron does not grow as large as quartz. As pure iron has a higher density, ultimately, momentum and impact energy of iron-rich aggregates are increased. As a consequence, iron-rich entities could fragment, which more likely would result in smaller clusters. However, this cannot account for a factor of 2 in size observed when adding only 25\,\% iron by mass and we consider this of minor importance here. Even without magnetic fields considered, the difference in sticking properties implies that clusters built from one or the other species have different sizes. We chose grain sizes for iron to be similar to the silicates as well as possible in order to compare the results more easily. Nature will not do so. Condensation will result in quite different dust sizes for different minerals and this will induce some size sorting going along with the size dependent sticking forces.

The tensile strength measurement indicates a drop of the effective surface energy of more than 2, comparing pure quartz and the sample with 50\,\% iron. This value characterizes sticking among dust grains composing a single aggregate. The ratio of the cluster sizes is different though, as it depends on sticking among the aggregates. The clustering could therefore be described by the concept of an effective aggregate surface energy.

Another obvious trend in Figure~\ref{fig.clustersize} correlates to the initial aggregate size set by the sieve. Smaller aggregates can be packed more tightly, which increases the number of contacts. Therefore, clusters composed of smaller aggregates are more stable and grow larger. However, we do not aim to quantify this effect here any further.

In the past, sticking properties have been considered for one given material \--- for example, quartz, water ice, or $\rm CO_2$ \citep{Blum2008, Gundlach2015, Musiolik2016} \--- and also the influence of an organic mantle on the collisional growth of silicates has been examined \citep{Homma2019}. Mixtures of different grains have actually never been studied in this context. What we find here, not related to iron and its magnetic properties, is that the differences of sticking properties of individual mineral grains will translate to different sizes of clusters depending on the fraction of each mineral species within an aggregate.

Overall, the size of clusters at the bouncing barrier is an important parameter since it plays a major role for the onset of drag instabilities (e.g. \citet{Drazkowska2014}). Furthermore, this might be important for a seemingly disconnected topic of chondrules. These (sub-)mm grains found in meteorites are sometimes highly size sorted \citep{Simon2018}. As small particles produce larger clusters, cluster growth is size selective and growth at the bouncing barrier might provide size sorting.

\subsection{Magnetic aggregation of iron-rich clusters}

\begin{figure}[h]
	\includegraphics[width=\columnwidth]{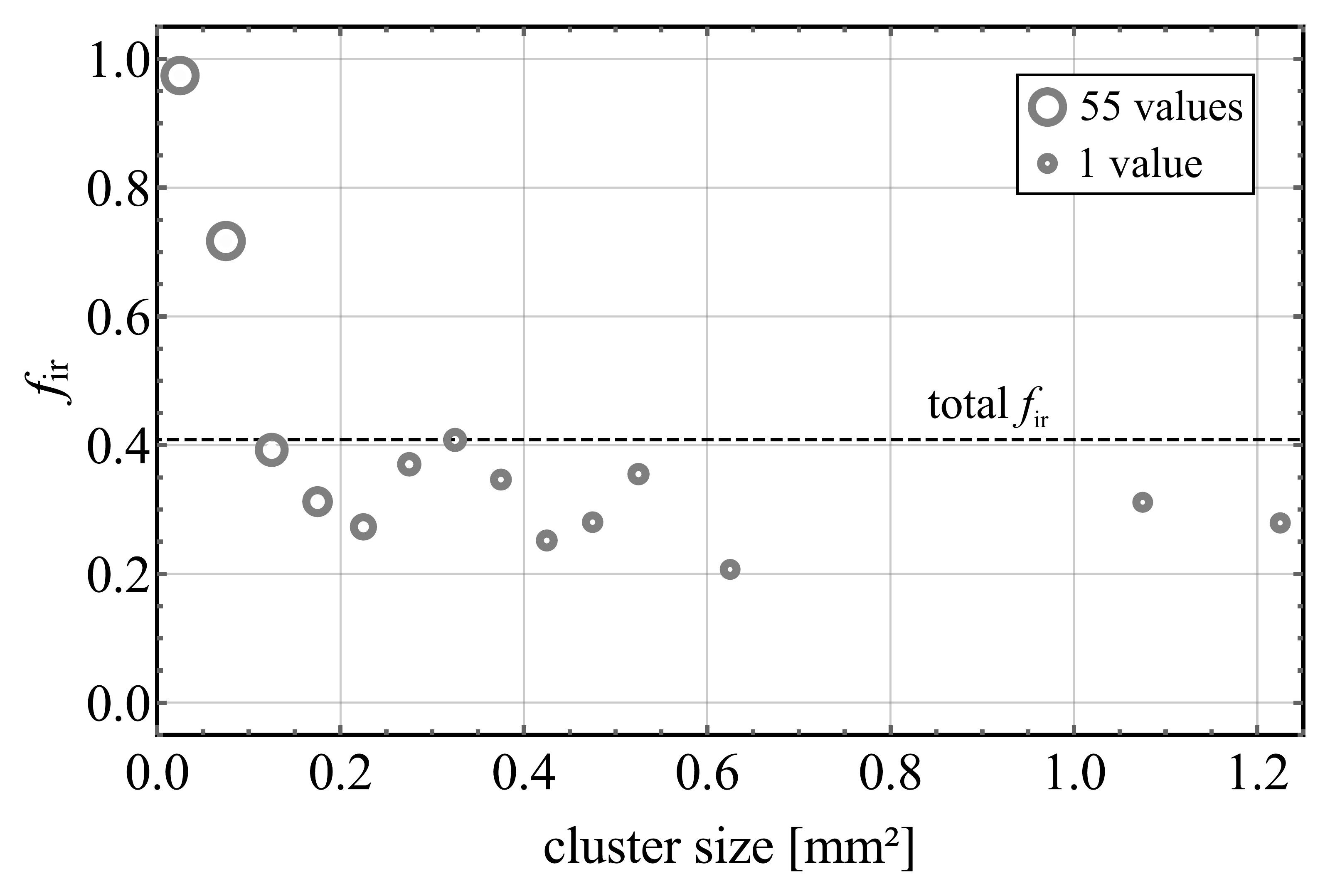}
	\caption{\label{fig.ironfraction1} Size of grown clusters and the area fraction of iron-rich material within these clusters without a magnetic field. The dashed line indicates the total area fraction of iron-rich material in the sample. As the data are binned (bin size 0.05\,mm$^2$), the size of the dots represents the number of data points in the respective intervals.}
\end{figure}

This study also aims at analyzing how a mixture of magnetic and nonmagnetic materials evolves, in particular, pure quartz and iron-rich aggregates with an iron mass fraction of 50\,\%. At first, the evolution of an ensemble of particles without magnetic field is discussed. Clusters rapidly reach a maximum size before collisions and gas drag destroy them again. There is a continuous fluctuation within the clusters, but the size distribution no longer changes. Figure~\ref{fig.ironfraction1} shows the area fraction of iron-rich aggregates $f_{\rm{ir}}$ in a cluster of a given size for one experimental run. The first observation reveals that the smallest clusters are iron-rich. Obviously, even though the experiment was seeded with both kind of aggregates being the same size, quartz aggregates grow larger while iron aggregates stay small, just as discussed in the previous section. Therefore, clusters can only grow bigger if a considerable amount of quartz is present.

Considering the difference in sticking properties of individual grains, this is actually not surprising. Necessarily, aggregates, being composed of one or the other species, have different sticking properties, too. Especially at the bouncing barrier, the kind of aggregate incorporated in the cluster is important for its stability. Due to the lower surface energy, clusters with iron-rich aggregates are less stable and fragment again more easily. Therefore, growth is biased toward the more sticky kind of grains.

The evolution of clusters is different with a magnetic field applied. As seen in Figure~\ref{fig.example2}, clusters are still mostly linear as in Paper I, since magnetic dipole-dipole interaction enhances sticking along the field lines. However, due to the nonmagnetic quartz aggregates, different cluster chains can be linked by quartz entities. Clusters can therefore grow somewhat larger as different strands are glued together. This is different from the observations in Paper I.

\begin{figure}[h]
	\includegraphics[width=\columnwidth]{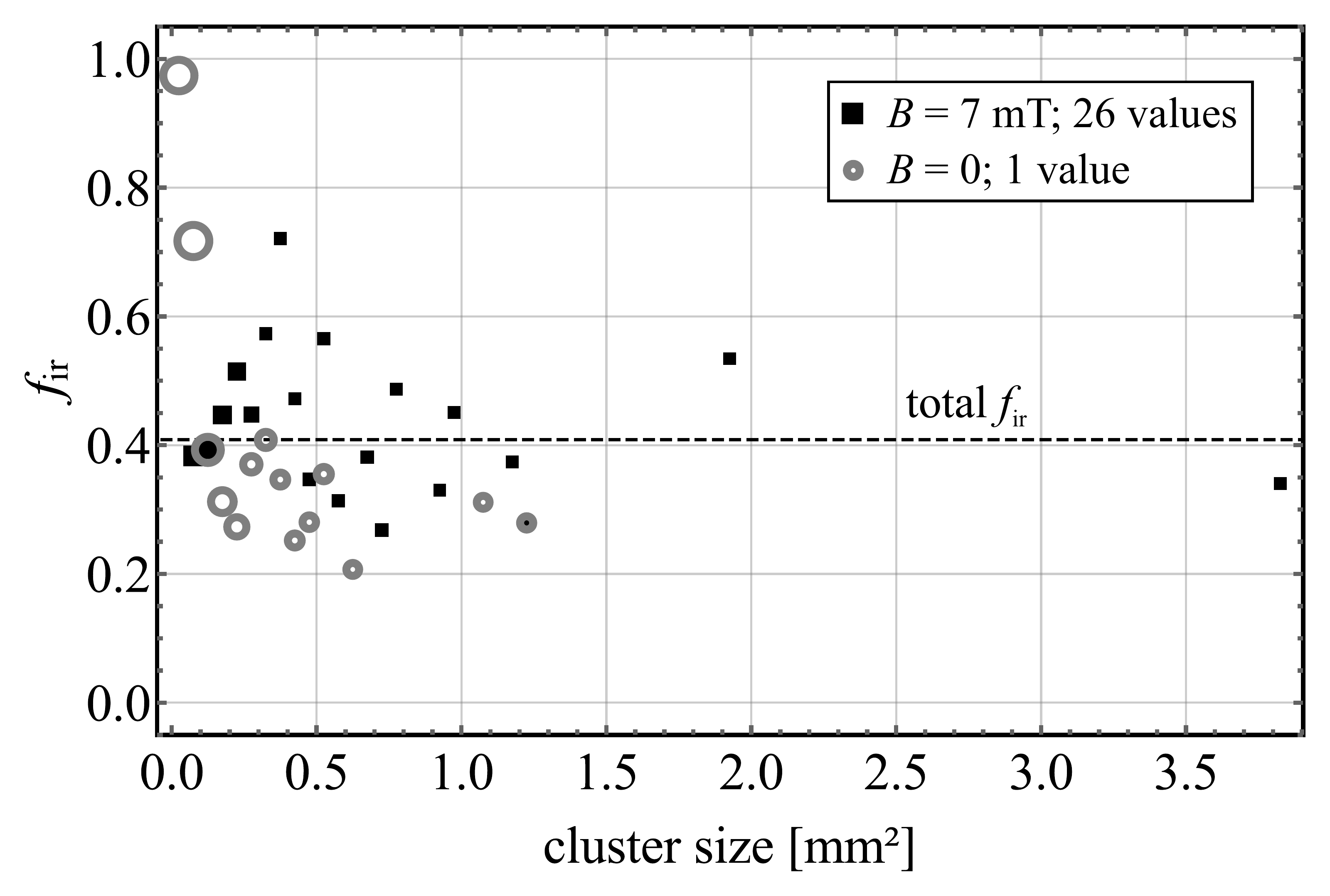}
	\caption{\label{fig.ironfraction2} Size of grown clusters and the area fraction of iron-rich material within these clusters with an applied magnetic field of 7\,mT. The size of the dots represents the number of data points in the respective intervals (bin size 0.05\,mm$^2$). The data without magnetic field are included in gray for comparison.}
\end{figure}

To quantify these observations, Figure~\ref{fig.ironfraction2} shows the area fraction of iron-rich aggregates $f_{\rm{ir}}$ within the magnetic field. There is a significant increase in the size of all clusters. The small iron-rich clusters decrease in numbers as they grow to larger entities due to magnetic boost. However, larger clusters are not restricted to iron-rich material. In fact, they contain significant amounts of quartz aggregates sprinkled over iron-rich chains and silicate clusters connecting the chains. Even larger clusters with less than the initial iron-to-quartz ratio form.

This boosted growth was observed for the ensemble with 40\,\% iron-rich aggregates by area as well as for a higher density of 70\,\%. At the magnetic field strength tested (7 mT), this effect is not very sensitive to the iron-to-quartz ratio of the total sample unless it is too low.

\begin{figure}[h]
	\includegraphics[width=\columnwidth]{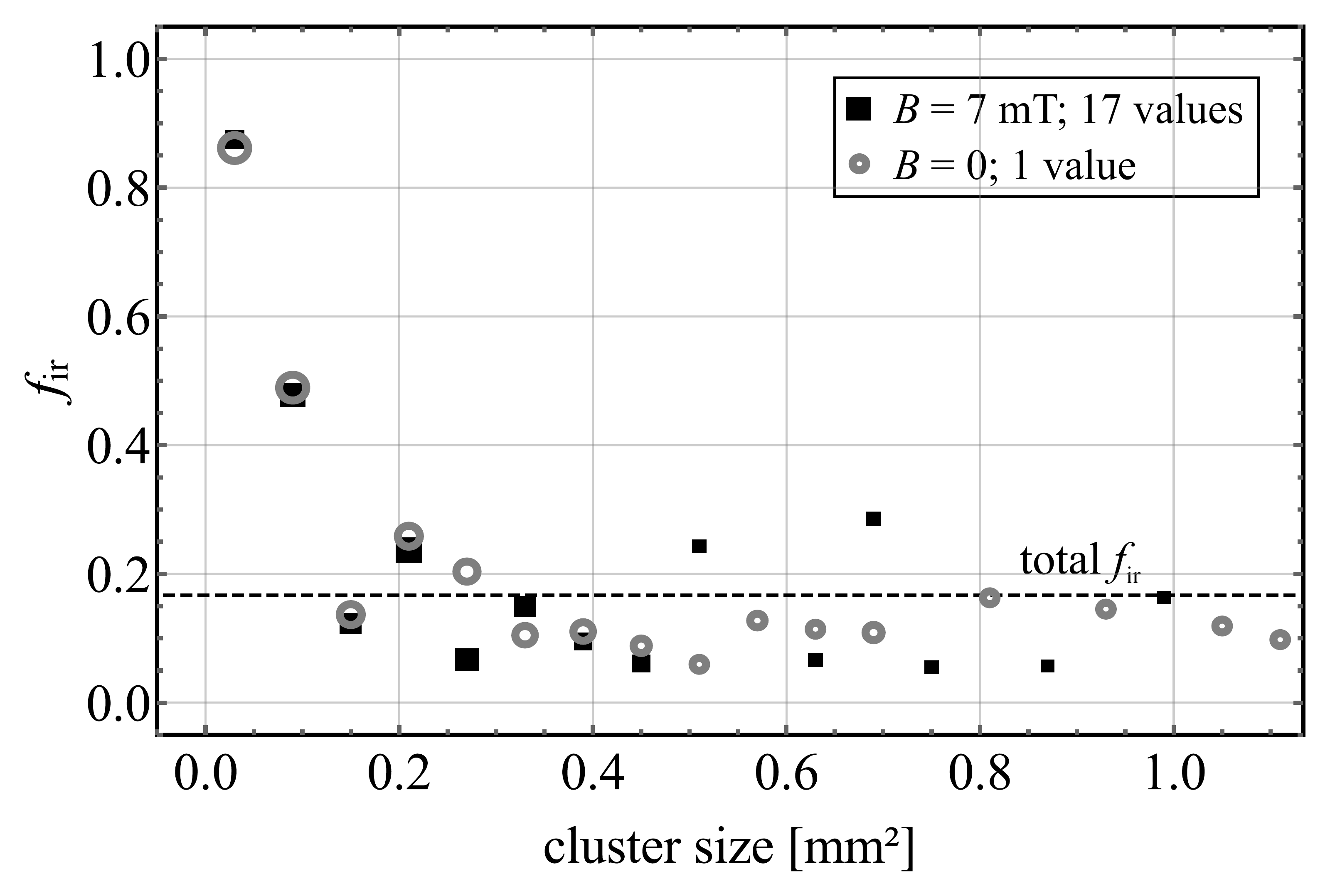}
	\caption{\label{fig.ironfraction3} Size of grown clusters and the area fraction of iron-rich material within these clusters. The total amount of iron-rich aggregates is not high enough for boosted growth.}
\end{figure}

In numbers, a too low area fraction of iron-rich aggregates in this context is about 20\,\%, which is shown in Figure~\ref{fig.ironfraction3}. In that case, magnetic fields no longer make a difference, while at higher iron fractions a boost of larger clusters was observed. Although we did not vary the iron fraction any further, this clearly indicates a threshold behavior. A few iron-rich aggregates start forming chains as the magnetic field is above the critical value for magnetic aggregation to occur as found in Paper I. However, they do not merge in one single chain, but they are mostly sprinkled over the larger quartz clusters that dominate the ensemble. We note that we neither varied the iron fraction within the iron-rich aggregates nor the magnetic field in this study. Both quantities together with the area fraction of iron-rich aggregates certainly influence whether growth of larger clusters is boosted or not.

\section{Conclusion}

There are a number of obvious and more subtle findings in this study. Foremost, in high magnetic fields, as found in parts of protoplanetary disks, clusters containing a fraction of iron grains grow larger than without magnetic fields as already revealed in Paper I. The addition of nonmagnetic aggregates does not change this earlier result fundamentally but modifies it in two ways. First, these aggregates can link different magnetic chains and make aggregates larger, as well as in the direction perpendicular to the magnetic field. Second, if more than 80\,\% of the aggregates are nonmagnetic, magnetic boost is suppressed. We did not vary the magnetic field strength in this study, but combined with the results from the earlier paper, there is still a bias in growth toward higher iron fractions. The threshold behavior suggests that favorable conditions are needed for magnetic boost. Specifically, a sufficient iron fraction has to be present, which is expected to be available in the inner disk. These results support the concept of magnetic aggregation that might help forming Mercury-like planets close to their host stars.

Somewhat unexpectedly, the iron-quartz system is but one example of a mixed sample where grains have different sticking properties. This is not unique to this system. There are natural differences in sticking properties of grains of different species. We speculate that this translates in a bias of cluster size for quite different mixtures, in general, if the constituents are a major component, silicate, iron, or ice.

Regarding the cluster size scale, the largest size at the bouncing barrier is a consequence of the properties of its constituents, i.e., the grain size, surface energy, and magnetic boost. Any size sensitive mechanism for further evolution  of the clusters, i.e. drag instabilities, will therefore lead to fractionation and sorting on a
still larger size scale.

\section{acknowledgements}

This project is funded by DFG grant WU~321/14-1. We also appreciate the very helpful reviews by Joseph Nuth and an anonymous referee.

\bibliography{bib}

\end{document}